\documentclass[useAMS,usenatbib]{mn2e}
\usepackage{amssymb}
\usepackage{graphicx}
\bibpunct{(}{)}{;}{a}{}{,} 

\def\kms{km ${\rm s}^{-1}$}
\def\sn1{${\rm s}^{-1}$}

\def\cmn{${\rm cm}^{-2}$}

\def\apm{APM 08279+5255}
\def\sp{\space}

\def\apj{{ApJ}}
\def\mnras{{MNRAS}}
\def\apjs{{ApJS}}
\def\pasj{{PASJ}}
\def\araa{{ARAA}}
\def\aap{{A\&A}}
\def\apjl{{ApJ}}

\title[low- and high-velocity outflows]{On the X-ray low- and high-velocity outflows in AGNs}

\author[Ram{\'i}rez and Tombesi]{J.M.~Ram{\'i}rez~$^{1}$\thanks{E-mail:jramirez@aip.de}
and F.~Tombesi~$^{2,3}$\\
$^{1}$Leibniz-Institut f\"{u}r Astrophysik Potsdam, An der Sternwarte 16 D-14482 Potsdam, Germany\\
$^{2}$Department of Astronomy and CRESST, University of Maryland, College Park, MD 20742, USA\\
$^{3}$Laboratory for High Energy Astrophysics, NASA/Goddard Space Flight Center, Greenbelt, MD 20771, USA}
\begin{document}
\bibliographystyle{mn2e}

\date{Accepted . Received ; in original form }

\pagerange{\pageref{firstpage}--\pageref{lastpage}} \pubyear{2011}

\maketitle

\label{firstpage}

\begin{abstract}
             An exploration of the relationship
             between bolometric luminosity and outflow velocity,
             for two classes of X-ray outflows in a large sample of
             active galactic nuclei has been performed.
             We find that line radiation pressure could be
             one physical mechanism that might accelerate the gas we observe
             in warm absorber, $v\sim 100-1000$ \kms,
             and on comparable but less stringent grounds the ultra-fast outflows (UFOs),
             $v\sim 0.03-0.3c$.
             If comparable with the escape velocity of the system;
             the first is naturally located at distances of the dusty torus, $\approx 1$ pc,
             and the second at sub-parsec scales, $\approx 0.01$ pc, in accordance with large set
             of observational evidence existing in the literature.
             The presentation of this relationship might give us key clues for our understanding of the different
             physical mechanisms acting in the center of galaxies, the feedback process and its impact on the evolution
             of the host galaxy.
\end{abstract}


%
\section{Introduction}

Mildly relativistic and 
non-relativistic absorption troughs 
are observed
in the X-ray spectra of 
active galactic nuclei (AGNs).
A qualitative separation is usually done
between classical $v\sim 100-1000$ \kms\sp warm absorbers,
we refer to here (throughout paper) as {\it low-velocity} outflows
\citep{blustin2005a,mckernan2007a},
and ultra-fast outflows (UFOs) $v> 10,000$ \kms,
we refer to here as {\it high-velocity} outflows
\citep{tombesi2010b,tombesi2210a}.

On one hand, 
the extracted velocity from the classical
$v\sim 100-1000$ \kms\sp
warm absorber is mainly obtained using the 
Fe M-shell  
$2p-3d$ unresolved transition array
(UTA) \citep{behar2001b,ramirez2008b},
and
O~{\sc vii} or O~{\sc viii} resonance lines
\citep[e.g.,][the hallmark of the classical warm absorber]{george1998a}.
The spatial location of this absorbing material
is uncertain. Models
of X-ray absorbers in AGN place them
at 
a wide range of distances from the central
source.
Specifically they are suggested to be
winds originating from the accretion
disk \citep{murray1995a,elvis2000a},
located at the dusty
($\sim 1$ pc) torus scales \citep{krolik2001a}
or even beyond the narrow-line
region \citep[e.g.][]{ogle2000a}.

On the other hand,
blueshifted Fe K absorption lines 
have been detected in
recent
years at $E>7$ keV
in the X-ray spectra of several radio-quiet AGNs
\citep[e.g.,][]{chartas2002a,chartas2003a,pounds2003a,dadina2005a,
markowitz2006a,braito2007a,ramirez2008a,cappi2009a,reeves2009a,
tombesi2010b,tombesi2210a}.
They are usually 
interpreted as
due to resonant absorption from Fe~{\sc xxv-xxvi}
associated with a zone of circumnuclear gas
photoionized by the central X-ray source, with ionization parameter 
$\log\xi\sim 3-6$ erg/s cm
and column density $N_H\sim 10^{22}-10^{24}$ \cmn. 
The energies of these absorption lines are systematically blueshifted 
and the corresponding velocities can reach up to mildly relativistic
values of $\sim 0.03c-0.3c$. These findings are important because they suggest the presence of 
previously unknown massive and highly ionized absorbing material outflowing from their nuclei,
possibly connected with accretion disk winds/ejecta 
\citep[e.g.,][]{king2003a,proga2004a,sim2008a,ohsuga2009a,king2010a,sim2010a}.

Several acceleration mechanisms have been proposed to explain these outflows:
(1) thermally accelerated winds \citep{krolik2001a};
(2) radiation pressure through Thomson scattering and magnetic forces \citep[MHD,][]{ohsuga2009a};
(3) and radiation pressure due to the absorption of spectral lines 
\citep[e.g.,][]{proga2004a,ramirez2008a,schurch2009a,sim2010b,ramirez2011a}.
Although the first one can explain the velocities we observe in the low-velocity
outflows, it can be 
excluded because it can not explain
the $\sim 0.1-0.2c$
we observe in UFOs \citep{tombesi2010b,tombesi2210a}.
\cite{ohsuga2009a}
seem to reproduce the velocities observed in {\it low-} and {\it high-velocity}
outflows.
On the other hand,
\cite{arav1994a,ramirez2008a,saez2009a,chartas2009a},
invoke radiation pressure due to lines to explain 
the $\sim 0.2c$
outflow detected in a good S/N X-ray spectrum of a
{\it high-z} quasar (the broad absorption line [BAL], \apm),
and they reproduce, as part of the procedure,
the Fe~{\sc xxv-xxvi} lines detected at $E>7$ keV.
Here we focus on this kind of approach, since
it allows us
to explore, very efficiently, a wide range of physical parameters
of the system. 

The goal of this Letter is to place a proof
of idea for a
systematic study
about the
operating
acceleration mechanisms in both;
X-ray {\it low-} and {\it high-velocity} outflows,
using an anisotropic radiative pressure framework
\citep[e.g.,][]{proga2000a,proga2004a,liu2011a},
beginning with line radiation pressure.

We present the observables from which we build the model in \S \ref{obser1}.
The details of the proposed model are presented in \S \ref{model1}.
The results and the discussion are in \S \ref{result1}.
We summarize in \S \ref{summ1}.
\section{Observables}\label{obser1}

In this section we present the observables of the two types of outflows,
since they are the initial motivation for the proposed model.

\subsection{The low-velocity outflows}
When describing the physical conditions of warm absorbers, it is common
to use the definition of ionization parameter
$\xi=\frac{4\pi F_{\rm ion}}{n_H}$
\citep{tarter1969a},
where $F_{\rm ion}$ is the total ionizing flux
($F_{\rm ion}=L_{\rm ion}/4\pi r^2$),
and $n_H$ is the gas density.
The source spectrum is described by the
spectral (specific, energy dependent $\epsilon$) luminosity
$L_{\epsilon}=L_{\rm ion}f_{\epsilon}$, where $L_{\rm ion}$ is the integrated
luminosity from 1 to 1000 Ryd, and $\int_{1}^{\rm 1000~Ryd} f_{\epsilon}
d\epsilon=1$.

So we describe the $\sim 1000$ \kms \sp warm absorber outflows
as absorbing material around a supermassive black hole (SMBH)
with mass
$M_{BH}\sim 3 \times 10^7$ \citep{peterson2004a,blustin2005a},
column density of the absorbing material
$N_H \sim 10^{20}-10^{22}$ \cmn,
flowing outwards at velocities $v\sim 100-2000$ \kms
\citep[e.g.,][]{kaspi2002a,krongold2003a,krongold2005a,ramirez2005a},
at medium ionization states
$\log \xi\sim 0-3$ erg s$^{-1}$ cm \citep{blustin2005a}.

When computing the energetics they find mass loss rates of
\footnote{Mean of the $\dot{M}_{out}$ reported by \cite{blustin2005a} in
their Table 4, excluding Ark 564 (outlier $\dot{M}_{out}=23$ M$_{\sun}$ yr$^{-1}$).}
$\dot{M}_{out}\approx 0.6$
M$_{\sun}$ yr$^{-1}$
ratios of $\dot{M}_{out}$ to accretion rates
\footnote{Mean of the $\dot{M}_{out}/\dot{M}_{acc}$ reported by \cite{blustin2005a} in
their Table 4, excluding Ark 564 (outlier $\dot{M}_{out}/\dot{M}_{acc}=550$).},
$\dot{M}_{out}/\dot{M}_{acc}\approx 5$
and
kinetic luminosity 
\footnote{Excluding PG 0844+349 and PG 1211+143.}
$L_{EK}=\frac{\dot{M}_{out}v^2}{2}$,
of the orders of $10^{38}-10^{41}$ erg/s,
representing less than 1 \% of the bolometric luminosity \citep{blustin2005a}.
The main conclusion from these estimations is that these outflows
contributes little to the energy injected
in to the host galaxy. But the amount of matter processed over the AGN lifetime
can be significant
\citep[also in accordance with][ for instance]{krongold2007a}.

\subsection{The high-velocity outflows}

The characteristics of the ultra-fast outflows with $v$$\ge$10,000 \kms \sp 
($\ge$0.03c) can be derived from 
the blue-shifted Fe~{\sc xxv-xxvi} absorption lines detected by \citet{tombesi2210a}
in a complete sample of local Seyfert galaxies. Such features are detected in $\sim$40--60\% of 
the sources, which suggests a covering fraction of $C$$\sim$0.5. 
Tombesi et al.~(2011) also performed a photo-ionization modeling of these lines.
They derived the 
distribution of the outflow velocities, which ranges from $\sim$0.03c up to mildly relativistic 
values of $\sim$0.3c, with a peak and mean value at $\sim$0.14c. 
As expected, these absorbers are highly ionized, with log$\xi$$\sim$3--6~erg~s$^{-1}$~cm, 
and have large column densities, in the range $N_H$$\sim$$10^{22}$--$10^{24}$~cm$^{-2}$. 

The SMBH masses of the Seyferts in the \citet{tombesi2210a} sample have a mean value of 
$M_{BH}=5.3 \times 10^7$~$M_{\sun}$
\citep{marchesini2004a,peterson2004a}.

When computing the energetics of these outflows, we see that these are more massive than 
the low-velocity ones, $\dot{M}_{out}$$\sim$ $0.1-1$~$M_{\sun}$ yr$^{-1}$ $\sim$$\dot{M}_{acc}$, 
and also much more powerful, with a mechanical power of $\sim$$10^{43}$--$10^{45}$~erg/s 
(e.g., Pounds et al. 2003; Markowitz et al. 2006; Braito et al. 2007; Cappi et al. 2009; Tombesi et al. 2010b). 
The latter value is $\sim$ $5-10$\% of the bolometric luminosity. 
Therefore, the high-velocity outflows may potentially play an important role on the 
expected cosmological feedback from AGNs \citep[e.g.,][]{king2010b,king2010a}.
\section{The Model}\label{model1}

We build our model based on observational evidence, basically from two sources:
\cite{blustin2005a} for the
classical $v\sim 100-1000$ \kms\sp warm absorbers,
referred here as {\it low-velocity} outflows,
and
\citet{tombesi2210a} for the
ultra-fast outflows (UFOs) $v\ge 10,000$ \kms,
referred here as {\it high-velocity} outflows.

As we see, together the two classes of outflows cover a wide 
range in velocity and black hole masses.
We seek a physical model which provides the context to explain 
(to first approximation),
part of the observational evidence we have up to now.

In order to do that, and to gain some insight into the relationship between outflow velocity
($v_{out}$) and luminosity, we invoke the velocity profile as a function of radius, given by
hydrodynamical calculations \citep{proga2000a}. The mathematical shape of the relation is given 
by assuming an outflow accelerated by radiation pressure from a central source with a bolometric
luminosity $L_{bol}$, and a mass of $M_{BH}$:

\begin{equation}
{ v_{out}[i] = \left [ 2GM_{BH}[i]\left( {\Gamma_{f}}{{L_{bol}[i]}\over{L_{Edd}[i]}} - 1\right) \left( {{1}\over{R_{in}}} - {{1}\over{R}} \right) \right]^{1/2}},
\label{luv1}
\end{equation}

\noindent
where $L_{Edd}$ is the Eddington luminosity, $\Gamma_{f}$ is the force multiplier,
where the acceleration due to the absorption of discrete lines is encapsulated
\citep{laor2002a},
$R_{in}$ is the radius at which the wind is launched from the disk, and $R$ is the distance of the
accelerated portion of the outflow from the central source.
The index {\it i}, runs
simultaneously over
the two distributions: velocity/BH mass.
Assuming that we observe the gas when it has reached
the terminal velocity of the
wind, i.e. $v_{out}$ at $R=\infty$, we can write:

\begin{equation}
L_{bol}[i]
= 
{{L_{Edd}[i]}\over{\Gamma_{f}}} \left({{v_{out}[i]^2R_{in}}\over{2GM_{BH}[i]}}+1 \right ).
\label{vout1e}
\end{equation}

Now we have the basic ingredient of the model, and we are ready to build our
two classes of outflows.

\noindent
{\it {\sc set~1}: High-velocity}. This synthetic sample is composed from 1000 black holes
with masses
normally distributed with mean, $\mu_m=2.6\times 10^7$~M$_{\sun}$ and standard deviation,
$\sigma_m=1.3 \times 10^7$~M$_{\sun}$
\footnote{
We take the middle point ($\mu_m=2.6\times 10^7$~M$_{\sun}$) between
the mean
extracted from \cite{tombesi2011a} ($\mu_m=5.3\times 10^7$~M$_{\sun}$),
excluding the mass of Mrk 205 (outlier $M_{BH}=44 \times 10^7$~M$_{\sun}$),
and
the mean
extracted from Table 5 of \cite{blustin2005a} ($\mu_m=2.7\times 10^7$~M$_{\sun}$),
excluding the mass of IRAS 13349+2438 (outlier $M_{BH}=80 \times 10^7$~M$_{\sun}$).
Also we take
the larger value of the observed dispersion $\sigma_m(obs)=1.3 \times 10^7$~M$_{\sun}$.
}.
Also we use a normal distribution for the outflow velocity
of the
absorbing gas around the BH with $\mu_v=57,000$ \kms
(this is the best-fit value for {\sc set~1}), and $\sigma_v=15,000$ \kms.

\noindent
{\it {\sc set~2}: Low-velocity}. In this case we use 1000 black holes
with masses
normally distributed with the same mean, $\mu_m=2.6\times 10^7$~M$_{\sun}$ 
and standard deviation, $\sigma_m=1.3 \times 10^7$~M$_{\sun}$
as before
but we use a normal distribution for the outflow velocity
of the
absorbing gas around the BH with $\mu_v=1800$ \kms
(this is the best-fit value for {\sc set~2}), and $\sigma_v=600$ \kms.

The best-fit values of $\mu_v(1)$ and $\mu_v(2)$ 
(as well $\Gamma_f(1)=250$ and $\Gamma_f(2)=40$,
see next section), are estimated by comparing a grid of models computed
using
$10,000 \leqslant \mu_v(1) \leqslant 70,000$ (\kms) and
$100 \leqslant \mu_v(2) \leqslant 2000$ (\kms), with
the corresponding set of observation:
{\sc set~3} {\it vs} \cite{tombesi2011a} and
{\sc set~4} {\it vs} \cite{blustin2005a}.
We performed the comparison using a Kolmogorov-Smirnov test.
We take the $\mu_v(1)$ and $\mu_v(2)$ 
(as well $\Gamma_f(1)=250$ and $\Gamma_f(2)=40$,
see next section) which give maximum p-values of the test.
The final p-value for the comparison {\sc set~3} {\it vs} \cite{tombesi2011a}
gives $\approx 0.8$  ($D=0.16$) and for the
{\sc set~4} {\it vs} \cite{blustin2005a} $\approx 0.07$ ($D = 0.38$).

The launching radius $R_{in}$, for each set of simulations is based on
the results of \cite{blustin2005a} for {\sc set~2}; i.e., $R_{in}({\sc set~2})=1~pc$
(orders of magnitude value). For {\sc set~1}, we set $R_{in}({\sc set~1})=0.01~pc$,
based on \citet{tombesi2010b,tombesi2210a}.


\begin{figure}
\includegraphics[angle=-90,width=75mm]{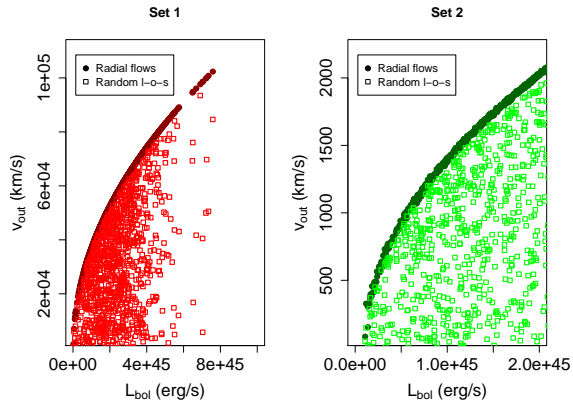}
 \caption{
Velocity of the outflow {\it vs}
the bolometric luminosity needed to accelerate the wind.
Filled circles are the velocities observed if we see
the wind along the flow. Open squares are including the effects
of random line-of-sights (see text for discussion).}
\label{vel1}
\end{figure}

\section{Results and Discussion}\label{result1}

In this section we discuss the results of our simulated samples and
summarize their  physical context and implications.

Figure \ref{vel1} shows the
theoretical relationship between outflow velocity 
and bolometric luminosity.

The curvature we see in $v_{out}$ of {\sc set~2} (and also {\sc set~1},
filled circles, our line of sight is along the flow, see below),
is due to 
the quadratic dependence of the luminosity
with $v_{out}$ given by equation \ref{vout1e}.
It is worth noticing that if we assume a force multiplier
of $\Gamma_f=40$ 
(this is the best-fit value for {\sc set~2} and {\sc set~4}, see section \ref{model1})
; we are able to reproduce the range
of luminosities seen in Figure 4 of \cite{blustin2005a}.

On the other hand this model is not reproducing well the
dispersion we observe in Figure (4) of \cite{blustin2005a},
which could be explained by the fact that
equation \ref{vout1e} will give radially accelerated flows, or
in other words, that we are observing all the objects radially
along the flow.
To include the effect of observing random line-of-sights
transversely through different sections of the flow we
connect the observed velocity $v_{obs}$ with the 
intrinsic radial velocity,
i.e. $v_{obs}=v_{out}\cos\theta$,
where $\theta$ is the angle between the outflow direction and the line of sight.
Then using 
the same luminosities produced by {\sc set~2},
but plotting (open squares) against 1000 $v_{obs}$(s) randomly computed
using a random generator of numbers (with $\pi/12\le \theta \le\pi/2$)
\footnote{The lower limit assumes that the torus cover $\sim 30$ degrees, 
so we are able to observe the outflow only from $\theta \ge \pi/12$.},
and equation \ref{vout1e} with $v_{obs}=v_{out}\cos\theta$,
we produce the sub-sample {\sc set~4}. It is easy to see that
{\sc set~4}, resembles better the plot shown in Figure 4
of \cite{blustin2005a},
taking into account the possible incompleteness of the sample.

\begin{figure}
\includegraphics[angle=-90,width=75mm]{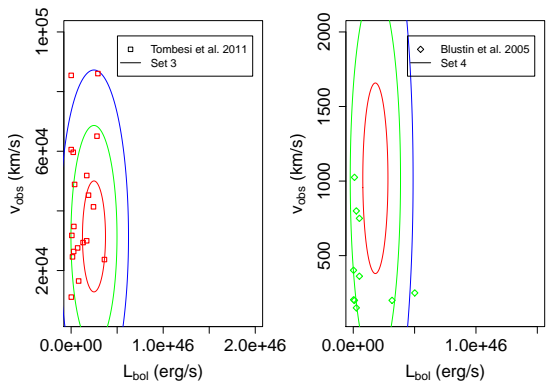}
 \caption{
Theoretical {\it vs} observational outflow velocity 
against
luminosity. Ellipses (1,2 and 3 $\sigma$ contours of the model)
are theoretical calculations coming from
two samples of outflows;
{\it Left panel:}
high-velocity ({\sc set~3}, i.e.,
including the effects
of random line-of-sights), and
{\it Right panel:}
low-velocity ({\sc set~4}, i.e., including the effects
of random line-of-sights). Open squares are observational points
from the XMM-Newton radio-quite sample of \citet{tombesi2011a},
for the UFOs.
The open diamonds are points compiled
by \citet{blustin2005a}.
}
 \label{vel2}
\end{figure}

Doing the same for the UFOs,
we use
the luminosities produced by {\sc set~1}
($\Gamma_f=250$, this is the best-fit value for {\sc set~1} and {\sc set~3}, see section \ref{model1})
but
plotting (open squares) against 1000 $v_{obs}$(s) randomly computed
using a random generator of numbers 
(with $\pi/12\le \theta \le\pi/2$),
and equation \ref{vout1e} with $v_{obs}=v_{out}\cos\theta$,
to produce the sub-sample {\sc set~3}. In this case
{\sc set~3}, better resembles the relationship between
$v_{out}$ and luminosity (see Figure \ref{vel2}).

In Figure \ref{vel2} we place both classes of outflows together along
with observational points taken from two samples
of objects:
14 points out of the 23 objects reported by \citet{blustin2005a}
(the others either did not report outflow velocity or were unknown),
in the right panel;
and 19 points from the sample of 19 objects were UFOs have been
detected by \citet{tombesi2011a}, in the left panel.

There are several interesting facts about the plot:
(1) {\it the observational points} cover $\approx 3.8$ orders of magnitude
in velocity;
(2) they cover $\approx 3$ orders of magnitude in bolometric luminosity 
(from $\sim 10^{43}-10^{46}$ erg/s);
and
(3) that our proposed model is able to reproduce 
most of the low-velocity points (11/14)
(two of the points might fall in the high-velocity set instead)
and, less stringently,
half of the points for the high-velocity set (11/19),
using one physical acceleration mechanism 
and three free parameters:
mass of the BH ($M_{BH}$),
outflow velocity ($v_{out}$)
and
force multiplier ($\Gamma_f$).
On the other hand, there are two sectors where the deviations
between model and theory are large:
(1) the high-velocity/low luminosity; and 
(2) the low-velocity/high luminosity, both
requiring a closer inspection to the completeness of the samples,
and/or the addition of other acceleration mechanisms,
like magnetic thrust for instance.
But, this may be the topic of future work.

\subsection{Anisotropic radiation pressure picture}

We propose here that the low- and high-velocity
outflows can be accelerated by the same
physical mechanism (as is suggested
by Figure \ref{vel2});
i.e., radiation pressure, and
that 
the differences depend on the values of the 3 fundamental 
parameters:
mass of the BH ($M_{BH}$),
the object luminosity ($L_{bol}$)
and
force multiplier ($\Gamma_f$).
{\it The values of $\Gamma_f$ we have used are of the
orders of those found in observational
\citep[e.g.,][]{laor2002a} and theoretical \citep[e.g.,][]{saez2011a} works}.
However, detailed photoionization
computations are required
to verify if the opacity of the gas under the physical conditions
presented here can overcome the over-ionization problem
\citep{proga2000a}, and are left for future work.

The anisotropic property of the radiation is basically
demanded by the existence of obscured and un-obscured
(Type 1 and Type 2) AGNs \citep{antonucci1993a},
and it explains
the decreasing of Type 2 AGNs as
a function of the X-ray Luminosity \citep{hasinger2008a}.
It is also intrinsically linked to the existence of the
dusty torus
\footnote{A convenient definition is that of cool ($\sim $1000 K)
optically and geometrically thick gas in approximate rotational
and virial equilibrium at $\sim$1 pc \citep{krolik1988a}.
}
and it gives the natural
frame to locate our two classes of outflows.
If comparable with the escape velocity of the system 
(a black hole of $M_{BH}=5.3\times 10^7$~M$_{\sun}$),
fast-outflow, $v_{out}\sim 0.10$c,
locates escape radius at $r\sim (20-100)r_S$
(Schwarzschild radius, $r_S=\frac{2GM_{BH}}{c^2}$).
This is very close to the
SMBH, and the origin is likely the accretion disk.

Again, using $500-1500$ \kms, as escape velocity from a
$2.6\times 10^7$~$M_{\sun}$ BH, locates the gas at $\sim 0.1-1$ pc,
well beyond the event horizon of the BH, and the 
broad line region \citep[BLR;][]{laor1993a}
as well,
where the dusty torus is thought to be located.
\section{Summary}\label{summ1}

For the first time,
an exploration on the relationship
between bolometric luminosity and outflow velocity,
for two classes of X-ray outflows in a large sample of
active galactic nuclei has been performed.
We find that:
(1) Line radiation pressure is an efficient
\footnote{Based on observed force multipler \citep{laor2002a}.}
mechanism to accelerate the low-velocity ($500-2000$ \kms)
gas we observe in the classical $\sim 1000$ \kms \sp
warm absorber.
(2) It might also become efficient
to accelerate the high-velocity ($0.05-0.3c$)
gas we observe in the UFOs.
(3) They both might be placed in the same context
of anisotropic radiation pressure 
\citep[e.g.,][]{proga2000a,proga2004a,liu2011a}.

However, there are many open questions, which will
require close investigation and detailed modeling.
In the first place the fact that we are assigning one type of 
outflow to different portions of the parameter space
($M_{BH}$, $L_{bol}$ and $\Gamma_f$),
does not preclude the existence of both in the same
object.
Also, careful studies
of
the connection between these outflows observed in X-rays
and outflows seen in
other bands of the electromagnetic spectrum,
UV, infrared or optical
(multi-wavelength studies) are important points
to be addressed.
Finally the close inspection of the main 
parameters of the system ($M_{BH}, v_{out}$ and $\Gamma_f$)
with cosmological parameters (like redshift) is of high relevance,
and might be the topic of future works.

\section*{Acknowledgments}

JMR would like to thank T. Kallman for a reading of the manuscript.
He also wants to thank the useful and constructive
comments from the referee which helped to improve several aspects of the work.
FT provided the data
based on observations obtained with the XMM-Newton satellite, an ESA funded mission with
contributions by ESA member states and USA. FT acknowledge support from NASA through
the ADAP/LTSA program. This research has made use of the NASA/IPAC Extragalactic
Database (NED) which is operated by the Jet Propulsion Laboratory, California Institute of
Technology, under contract with the National Aeronautics and Space Administration.



\end{document}